# Application of a dynamical two-box surface-atmosphere model to the Mount Pinatubo cooling event

Robert S. Knox and Kevin J. LaTourette*
Department of Physics and Astronomy
University of Rochester, Rochester, NY 14627-0171

**Abstract**

We analyze the global temperature change due to the Mt. Pinatubo eruption using a simple two-layer model of the atmosphere and surface to obtain results consistent with satellite data. Through analytic and numerical analysis we find a principal characteristic response time of 5 to 8 months and a climate sensitivity of 0.17 to 0.20 C/(W/m^2), corresponding to a negative instantaneous feedback. Our solutions were fit to the data, reproducing the results of a one-box model, and providing somewhat more detailed information about the feedbacks related to surface layer temperature. The formalism for coupling of the surface layer to the thermocline is set up but not applied.

## 1. Introduction

A leading goal of climatology is to determine the effect of various forcings on Earth's climate system. A forcing is defined as the effective change of net radiative input at the top of the atmosphere, and has been adapted to all forms of disturbance [1]. We have a particular interest in the eruption of Mount Pinatubo because the event provides a substantially greater disruption to the climate when compared with other climatic disturbances. With its climax occurring June 15, 1991, the eruption created the greatest aerosol cloud of the 20th century, releasing sufficient $SO_2$ into the atmosphere to reduce the average global temperature briefly by approximately 0.5° C. This powerful climatic event had the potential to "…[exceed] the accumulated forcing due to all anthropogenic greenhouse gasses added to the atmosphere since the industrial revolution began." [2] Occurring over a period relatively free of other climate forcings including the ENSO (El Niño-Southern Oscillation) effects and solar fluctuations, the Pinatubo event is an excellent candidate for analysis.

---

*Present address: Department of Applied Mathematics, University of Arizona, Tucson, AZ 85721.
 Email: rsk@pas.rochester.edu, klatourette@math.arizona.edu

Lindzen [3,4] studied volcano temperature anomalies by setting up a one-box model of the climate system with coupling to the thermocline. (The thermocline is an ocean layer in which temperature drops from its surface value. A layer of varying depth, of the order of dozens of meters, called the mixing layer, lies between the actual surface and the thermocline. The mixing layer temperature is uniform, as a result of turbulence caused by surface winds.) The forcing used was appropriate to that due to the 19th-century Krakatoa eruption, and means for approximating the exact solution were taken, resulting in a prediction of the correct order of magnitude of global cooling, but with a fairly poor fit of the result to the data. It has been argued [5] that the reason for the poor fit was that the characteristic relaxation time $\tau$ in the box was assumed to be of the order of several years, rather than several months as the data imply. The resulting formalism implied a climate sensitivity of $\lambda = 0.4$ C/(W/m$^2$) (to be discussed below, sec. IV).

Douglass and Knox [5] applied the Lindzen theory, the forcing calculations of Hansen [6,7], and the aerosol optical density estimates of Ammann *et al.* [8] to the Pinatubo case, producing a rather good fit to the data with a concomitant smaller $\tau$ of 6.8 months. They neglected the coupling to the thermocline. The climate sensitivity [1] predicted was $\lambda = 0.18$ C/(W/m$^2$), implying negative feedback, in apparent disagreement with the prevailing thinking about climate response. As a result, two comments were published [9,10] whose salient claim was that the neglect of coupling to the thermocline had caused the work of [5] to be invalid, because its characteristic decay time was too short. The authors responded [11,12] with a revised calculation that included a crude estimated coupling, finding $\tau = 5.8$ months, $\lambda = 0.18$ C/(W/m$^2$).

Being concerned about the approximations involved in the thermocline coupling, Douglass *et al.* [13] solved the surface-thermocline coupling problem exactly. This introduces one new important physical parameter, namely, the eddy diffusion constant $\kappa$. They found results very similar to those of [5,11,12], but the value of $\kappa$ required to fit the data was quite small, $2 \times 10^{-6}$ m$^2$/s, as compared with the usually assumed value of $1 \times 10^{-4}$ m$^2$/s.

A one-box model for the surface and atmosphere is a serious oversimplification, as there are obviously many physically different layers and the interactions between these layers cannot be accounted for. Splitting the box into a surface layer and atmospheric layer [14,15] allows us to account for the different fluxes that occur between the two. The one-box model also lacks a temperature structure in the atmosphere, preventing it from handling the distinct radiative flows at the top and bottom of the atmosphere layer. Differences in the radiative flow at the top and bottom are related to the temperature differences, and our particular two-box model includes this climatic feature [14] without significantly increasing the complexity of the model. Adding a third box to the model will allow for the inclusion of the deep ocean (thermocline) coupled with the surface layer and provides an exact mathematical solution. In this preliminary report we develop the formalism for the thermocline coupling but do not proceed to the necessary numerical computations. Thus, our two-box numerical results are to be compared with the results of the first Pinatubo calculation [5].

In Section II we describe our dynamical model for a generalized forcing. In Section III we apply the Pinatubo forcing, in Section IV climate sensitivity and feedbacks are discussed, and finally Section V is a summary.

## II. The dynamical model

We begin with a static model [14] for globally averaged quantities. This will enable us to establish values of some important parameters. Our model has an incoming solar radiation, $S_0 = 342$ W/m$^2$, averaged over the surface of Earth, and proposes the following energy balance equations between the two layers in the steady state (see [14] and Figure 1):

$$(1+b)\varepsilon F_A - \varepsilon F_S = AS_0 + S_{NR} \qquad (1)$$

and

$$-b\varepsilon F_A + F_S = BS_0 - S_{NR} - Q, \qquad (2)$$

representing the atmosphere and surface layers, respectively. $F_A$ and $F_S$ are the ideal Stefan-Boltzmann fluxes given respectively by $\sigma T_A^4$ and $\sigma T_S^4$, and $\sigma = 5.67 \times 10^{-8}$ W/m$^2$/K$^4$. The long-wave imbalance parameter $b$ is the ratio of the net downward flux at the lowest portion of the atmosphere to the net upward flux at the top of the layer. This acts as a proxy for the lapse rate (temperature gradient), which is negative with increasing altitude, leading to an imbalance of long-wave radiation. The temperature at the bottom of the atmosphere is $b^{1/4} T_A$. The long-wave emissivity $\varepsilon$ also acts as absorptivity in the term $\varepsilon F_S$ through Kirchhoff's Law. Parameters $A$ and $B$ are the fractions of incoming solar radiation ($S_0$) absorbed by the atmosphere and surface, respectively. Imbedded within $A$ and $B$ are corrections for the multiple reflections between the two layers, but $A$ and $B$ may also be understood as independent adjustable parameters under the condition that the overall system reflectivity (planetary albedo) is $\alpha = 1 - (A+B)$ [14]. The non-radiative flux, $S_{NR}$, is a result of turbulent exchanges of heat and latent heat between the surface and atmosphere, and represents a substantial fraction of the up-flow from the surface (approximately 100 W/m$^2$). We also wish to incorporate feedbacks, and will handle them in detail below. $Q$ represents the net heat flux from the surface to the deep ocean at the top of the thermocline. It is assumed zero in the steady state.

All parameter values are chosen so that $T_{S0} = 288K$ and $\alpha \approx 0.30$, both of which are accepted values. This forces the parameters into a limited possible range and ensures that their values reflect their natural states. Without these constraints, a generalization of the energy balance model would be unrealistic as it would not necessarily be even a crude representation of Earth's climate system.

We now make this model time dependent by writing (see Appendix A):

$$c_A \frac{du_A}{dt} + K_{AA} u_A + K_{AS} u_S = \Delta[A_0(t) S_0(t)] \tag{3}$$

and

$$c_S \frac{du_S}{dt} + K_{SA} u_A + K_{SS} u_S = \Delta[B_0(t) S_0(t)] - Q(t), \tag{4}$$

where $u_A = \Delta T_A$ and $u_S = \Delta T_S$ are anomalies, *i. e.*, temperature deviations from steady-state values. $c_A$ and $c_S$ are, respectively, the effective area specific heats of the atmosphere and surface boxes. The relaxation and coupling factors $K_{IJ}$, defined in Eqs. (A14), result from expanding Eqs. (1) and (2) in Taylor series, dropping all but terms linear in the temperature anomalies and the forcing variations $\Delta A_0 S_0$ and $\Delta B_0 S_0$. In our application of the model, subscript "$A$" refers to that part of the atmosphere above 2 km, containing about 76% of its mass, and "$S$" refers to the surface plus the lower 2 km of atmosphere. Thus $c_A = 0.76 \times 1.02 \times 10^7$ J/m$^2$/K and $c_S$ is an effective specific heat that includes the lower atmosphere contribution along with that of the actual surface, or $c_S = 0.24 \times 1.02 \times 10^7$ J/m$^3$/K + $h_m C_S$. Here $h_m$ is an effective mixed-layer depth [13] and $C_S$ = 4.1 x $10^6$ J/m$^3$/K, the volume specific heat of seawater. We have defined the "A" layer in this manner because the temperature data available refers to satellite measurement at roughly 2 km, and it can therefore be represented by $b^{1/4} T_A$. The addition of a third box

-- 6 --allows for the inclusion of the thermocline into our model (see Figure 1), and is accounted for using the method of *Douglass et al*. [13]. The flux at the interface of the thermocline and surface layer is $Q = -C_V \kappa (\partial u_T / \partial x)_{x=0}$, where the thermocline anomaly $u_T$ varies with depth $x$, measured downward. Here $\kappa$ is the effective eddy diffusion coefficient and the specific heat $C_V$ is the volume specific heat of sea water (the coupling at the thermocline interface, which is below the mixed layer at a depth of dozens of meters, does not involve the complication of land and atmosphere). Under Laplace Transformation with Laplace variable $p$, Eqs. (3,4) become

$$(c_A p + K_{AA})\tilde{u}_A(p) + K_{AS}\tilde{u}_S(p) = \mathcal{L}\{\Delta[A_0(t)S_0(t)]\} \tag{5}$$

and

$$K_{SA}\tilde{u}_A(p) + \left(c_S p + K_{SS} + C_v \kappa^{1/2} p^{1/2}\right)\tilde{u}_S(p) = \mathcal{L}\{\Delta[B_0(t)S_0(t)]\}, \tag{6}$$

where $\mathcal{L}\{...\}$ indicates LaPlace transform. In the transforms we have assumed that solutions of interest are zero at $t = 0$. Here, we will treat the case of constant solar flux, so the driving terms become $S_0 \tilde{A}(p)$ and $S_0 \tilde{B}(p)$. Note that the thermocline effect, represented by the term containing $\kappa$, appears in the surface equation. This term includes a factor $(p/\kappa)^{1/2}$ that originates in the factor $(\partial \tilde{u}_T(x,p)/\partial x)_{x=0}$ of $Q$. The thermocline anomaly $u_T$ itself has disappeared from the equations because its relevant value, $u_T(x = 0)$, is equal to $u_S$.

## III. Application to Pinatubo

### A. Case of volcano forcing

As in the earlier work [5], we take the quantitative effect of the Pinatubo event as a globally averaged reduction of solar intensity caused by an increased aerosol reflectivity, resulting in a forcing that is proportional to the increased aerosol optical density (AOD). This is represented in our model by

$$\Delta F(t) = -0.439 A(t/t_V) \exp(-t/t_V) \quad \text{W/m}^2, \tag{7}$$

where $A$ has been theoretically derived [6,7], and $t_V$ is the time of peak AOD (7.6 months after $t = 0$, the beginning of the event). For historical reasons and without danger of confusion we retain "$A$" as the constant on the right side of (7), although it is also our symbol for the atmospheric fraction of solar absorption. For its value, we use the most recent calculated by Hansen and coauthors [7], $A = 21$. With the notation $k_V = 1/t_V$, the Laplace transform of this forcing is

$$\Delta \tilde{F}(p) = -0.439 A k_V (p + k_V)^{-2}. \tag{8}$$

The forcing under consideration originates in the excess reflection due to aerosols and is expressed in the standard way in terms of flux change at the top of the atmosphere. This does not necessarily mean that it is the atmosphere layer that receives the full effect of the forcing. We assume that the effect will be felt in proportion to the absorption of solar energy by each of the layers, i. e., a fraction $\phi_A$ is assigned to the atmosphere layer and $\phi_S$ to the surface layer, with $\phi_A + \phi_S = 1$. The fractions will be in the ratio $\phi_A / \phi_S = A/B$, where $A$ and $B$ are the static model parameters (see eqs. 1 and 2 and Fig. 1). Thus the forcings to appear in eqs. (5,6) – replacing the entire right hand sides – are $\Delta \tilde{A}_0(p) = \phi_A \Delta \tilde{F}(p)$ and $\Delta \tilde{B}_0(p) = \phi_S \Delta \tilde{F}(p)$.

With these forcings and with the notation $k_{IJ} = K_{IJ}/c_I$, Eqs. (5) and (6) take the compact form

$$(p + k_{AA})\tilde{u}_A(p) + k_{AS}\tilde{u}_S(p) = (\phi_A/c_A)\Delta\tilde{F}(p) \tag{9}$$

and

$$k_{SA}\tilde{u}_A(p) + \left(p + k_{SS} + \kappa'^{1/2} p^{1/2}\right)\tilde{u}_S(p) = (\phi_S/c_S)\Delta\tilde{F}(p). \tag{10}$$

The additional abbreviation $\kappa'^{1/2} = \kappa^{1/2}(C_V/c_S)$ with dimension s$^{-1/2}$ has been introduced.

In the case $\kappa' = 0$ the functions $u_A(t)$ and $u_B(t)$ are elementary linear combinations of exponential functions. Otherwise a numerical attack on the inverse transform must be mounted. For the applications in this report, confined to the case $\kappa' = 0$, the exponential solutions have been obtained and are set out in Appendix B.

**B. Parameter determination**

Let us review the several parameters involved in fitting the volcano-induced temperature data with our model. The eigenvalues $\lambda_1$ and $\lambda_2$ (Eq. B2) are obtained from the $k$'s of Eq. (B1), which come from the area specific heats and the kinetic coefficients given in Eqs. (A7). Because of the fairly large number of parameters, we must adopt a systematic approach. One subset of parameters, called "static," is set by reference to steady-state properties. The static set consists of $\{S_0, b, \varepsilon, A, B, S_{NR}\}$, which may be seen in Eqs. (1,2), Fig.1 and Table 1. These parameters determine $F_A$ and $F_S$ or equivalently $T_A$ and $T_S$. As explained in [14,18], a reasonable static set is chosen by imposing several conditions: $T_A = 288$ K, albedo in the range 0.30-0.33, $S_{NR}$ approximately 100 W/m$^2$, $\varepsilon$ in the range 0.8-0.9, and a value of $b$ that produces a lower-atmosphere temperature consistent with observed lapse rates in conjunction with the predicted $T_A$. The solar constant is always taken to be 342 W/m$^2$. Since there are many more static parameters than associated observables, we must be prepared to test the sensitivity (or "robustness") of any subsequent results to the particular set of static parameter values. Here we do this by running the entire calculation for two distinct static sets (see Table 1).

From Eqs. (A14-16) we see that of the static quantities only parameters $\{b, \varepsilon\}$ and quantities $\{T_A, T_S\}$ will be needed in the dynamic problem. Here $T_A$ and $T_S$ determine $q_A = 4\sigma T_A^3$ and $q_S = 4\sigma T_S^3$. Since $T_S$ is fixed at 288 K, we have $q_S = 5.42$ W/m$^2$K. The other three quantities have limited ranges within the criterion of reasonableness. For static set 1, we have centered our dynamical calculations on the values $b = 1.65$ and $\varepsilon = 0.786$. Whenever these were changed, the others were adjusted to maintain $T_S = 288$ K and the satisfaction of other static criteria. These adjustment covered ranges of $1.55 \leq b \leq 1.80$ and $0.786 \leq \varepsilon \leq 0.810$ and did show that the results of the dynamical parameter fit were robust. Static set 2 differs considerably and still yields similar dynamical results.

Returning to the fits of time-dependent data, we define the "dynamical" set of parameters as $\{c_A, c_S, [f]\}$, where $[f]$ refers to the set of four feedbacks in Eqs. (A7). Because of our definition of the model and the atmosphere and surface layers (see Sec. II), $c_A$ is fixed and $c_S$ depends only on $h_m$, an effective ocean mixing-layer depth. The set of dynamical parameters is thus further reduced to $\{h_m, [f]\}$. It is clearly futile to vary all four feedbacks with a single data set under analysis, so we arbitrarily focus on $f_{SS}$ and $f_{AS}$, namely, those induced by surface temperature changes.

**C. Data Fitting**

Our primary data source is the global monthly satellite Microwave Sounding Unit lower troposphere temperature (TLT) anomaly data set [16], modified with El Niño and solar irradiance cycle perturbations effectively removed (see the discussion in [5], section 2.2). All reported analysis is based on this modified data set, denoted TLTm. Furthermore, because of the noise obviously existing in the set at times beyond $8.4\, t_V$, only the 61 points between $t = 0$ and $8.4\, t_V$ were used in the data fits (see any of Figs. 2-4).

Previously, TLTm has been regarded as the temperature change at the undifferentiated "surface" layer, although measurement takes place in the lower levels of the atmosphere. The present model allows us to be a bit more precise about this, and so, as discussed above, we fit the quantity $b^{1/4} u_A$ to the data. In what follows, this quantity will be referred to as $u_B$.

*1. Static parameter set 1*

The fitting began with all feedbacks assumed to be zero. In this case, it being the only other variable parameter in our protocol, $h$ was varied, starting with the arbitrary value 15 m, until the sum of the quantity $(u - \text{TLTm})^2$ was minimized. This least-squares method produced the value $h_m = 15$ m and an $R^2 = 0.71$ (when the fit was $u = u_S$) and 0.69 (when the fit was to $u = b^{1/4} u_A$). Values of $h_m$ as high as 30 m were considered. We note that the eigenvalues of the kinetic matrix at this stage were $9 \times 10^{-7}$ s$^{-1}$ and $4.3 \times 10^{-8}$ s$^{-1}$, corresponding to lifetime parameters of $0.056 t_V$ (13 days) and $1.17 t_V$ (8.9 months). From the solutions we can see that the larger eigenvalue (shorter lifetime) tends to contribute a negligible amount in the earlier part of the TLTm time course (see any of Figs. 2-4).

To next refine the fit, the feedback $f_{SS}$ was introduced and treated in the same way, minimizing the sum of squares and maximizing $R^2$, holding $h$ constant at its previously determined value. The result was a visibly better fit. This process was repeated with $f_{AS}$ only, then with both feedbacks, switching back and forth until a maximum $R^2$ was obtained. The resulting best fits, both with $R^2 = 0.74$, are shown in Figs. 3-4. ($h_m$ was again varied between alternation of feedbacks, with no changes in its required value.) It was not possible to assign a "better fit" to either $u_B$ (Fig. 3) or $u_S$ (Fig. 4), since both produced the same value of $R^2$. As explained earlier, we consider the $u_B$ fit to be the more realistic because of the nature of the data. Clearly, Fig. 4 shows that *either* of the solution sets is adequate to explain the data for the parameters that deliberately make $u_B$ fit.

*Value of the mixed layer depth.* As we adjust the values of the mixed layer $h_m$, both temperature minima ($u_S$ and $u_B$) occur later as $h_m$ increases, but only slightly. The two most notable changes induced by modifying $h_m$ are in the amplitude of the minimum temperature and the characteristic response time. By doubling $h_m$ to a value of 30 m, we see the amplitude decrease by $0.1 \pm 0.05$ C, and the climate response time doubles as $h_m$ doubles. When a value significantly greater than 20 m is used for the mixed layer depth, however, the response time does not follow the trend of the data, growing to the order of years not months. The best value for $h_m$ was $17.0 \pm 4.0$ m.

Varying the static parameters to check the robustness of the steady-state model made only slight changes in the value of $R^2$ (of order 0.0005). Finally, we note that all fits were done with surface forcing ($\phi_S = 0.97$, $\phi_A = 0.03$). There was virtually no effect of small variations of these fractions.

Results are shown in Table 2. Of particular interest are the feedbacks, $-0.31$ and $-0.62$, to be discussed in the section IV.

### 2. Static parameter set 2.

Parameter set 2 differs from set 1 principally in that a larger fraction of the solar absorption is allotted to the atmosphere. As a result, somewhat different values of the other parameters are required to produce the usual surface temperature and albedo, and the temperature of the atmosphere layer is increased. Within the philosophy of using elementary climate models, these parameter variations are acceptable if further results based on them are essentially unchanged. We find this to be the case here, in that nearly the same dynamical parameters are required to fit the Pinatubo data as well as does parameter set 1.

The best fit with parameter set 2 was found to be that for the "bottom temperature" $b^{1/2}u_A$. The solutions fitted were virtually identical to those of parameter set 1 and are not shown here. The 16% larger value of $h_m$, therefore the effective heat capacity of the surface box, probably reflects the need for "slowing down" the kinetics during the rising phase of the signal, since the increased involvement of the atmosphere brings in the effect of its smaller heat capacity at earlier times.

## IV. Climate sensitivity and feedback

The present analysis, to the extent that it is successful, determines the feedback parameters required in computing the parameter called sensitivity, defined as the steady-state temperature shift per unit step-function forcing, $\lambda = \Delta T_S/\Delta F$. For purely solar forcing, the no-feedback sensitivity for our case is given by [17]

$$\lambda_{S0} = \frac{\Delta T_S}{\Delta F} = \frac{\Delta T_S}{(1-\alpha)\Delta S_0} = \frac{T_{S0}}{4(1-\alpha)S_0} \cdot \frac{1}{1-\gamma}, \text{ where } \gamma = \frac{S_{NR}}{[(1+b)+bA]S_0}. \quad (13)$$

The sensitivity in the presence of feedback is $\lambda_S = \lambda_{S0}/(1 - f^S_{\text{eff}})$, where $f^S_{\text{eff}}$ is a straightforward but rather complicated mixture of the feedbacks (Appendix A, Eq. A18). The value of $\lambda_{S0}$ is commonly considered to be 0.30 °C/(W/m$^2$) [1] but has been corrected to 0.36 within the context of the present model [14,17].

The dynamics of the one-layer Pinatubo treatment [5] contained a single feedback parameter $f$ that can be identified with our $f_{\text{eff}}$. A negative effective feedback was indicated in that work, and, as seen above, we also require negative feedbacks. Any attempt to connect the value found in [5] to our model introduces many new undetermined parameters, as seen in Appendix A. Therefore, as an exploratory example, let us assume that the temperature dependence of non-radiative transfer, represented by $q_{NR} = dS_{NR}/dT_S$, is a dominant feedback cause, to the exclusion of all others. It is not hard to imagine that the non-radiative flux will increase with surface temperature, and therefore the fitting-predicted negative feedbacks are consistent with our model. According to Eqs. C15 and C17, $f_{AS} = -q_{NR}/\varepsilon_0 q_S$ and $f_{SS} = -q_{NR}/q_S$. These equations predict 2.0 and 1.43 W/m$^2$/K, respectively, for the value of $q_{NR}$. Comparison may be made with two estimates in the literature. The two-layer model work of Barker and Ross [19] predicts $q_{NR} = 0.33$ W/m$^2$/K. The thermodynamic treatment of Hartmann [20] predicts that the "sensible heat" part of $q_{NR}$ will be 12 W/m$^2$/K. This wide disparity calls for much more careful analysis of the model and the meaning of $S_{NR}$.

For the non-zero feedbacks found here, we have (Eq. C20)

$$f_{\text{eff}}^S = \frac{(1+b)f_{SS} - b\varepsilon f_{AS}}{1+b-b\varepsilon}. \qquad (14)$$

Using the static parameters employed in obtaining the feedbacks, one predicts $f^S{}_{\text{eff}}$ = –0.93 (static set 1) and –1.02 (static set 2). These are in remarkably good agreement with the value –1.0 ± 0.4 found in the earlier work [5].

Looking further into the feedback signs, it can also be argued that the term $B_1=(\partial B/\partial T_S)_0$, which is a positive-signed component of $f_{SS}$, also has a negative value, as follows: as the surface temperature increases, cloud cover will also increase, which will raise the absorption $A$ in the atmosphere and increase reflectivity, both of which will tend to decrease the absorption $B$ of the surface layer.

## V. Summary/Conclusion

Our solutions of the dynamical two-box model applied to the Mt. Pinatubo effect are entirely consistent with earlier work that modeled the system with one box. The parameter fits, while necessarily very crude because of the lack of multiple data sets and the number of parameters required, pass tests of reasonableness and provide more detailed information about surface-temperature-induced feedbacks. These feedbacks are introduced here by reference to temperature dependence of parameters in a two-box surface-atmosphere system. We believe that this treatment is new and can increase the applicability and value of the simple model.

## Appendix A. Basis of the time dependent model

Equations (1) and (2) are generalized to the time-dependent case as follows:

$$\frac{dE_A}{dt} = c_A \frac{d\Delta T_A}{dt} = -(1+b)\varepsilon F_A + \varepsilon F_S + AS_0 + S_{NR} \tag{A1}$$

and

$$\frac{dE_S}{dt} = c_S \frac{d\Delta T_S}{dt} = b\varepsilon F_A - F_S + BS_0 - S_{NR} - Q, \tag{A2}$$

where $E_A$ and $E_S$ are the energy content per unit area of the $A$ and $S$ boxes, respectively, and $c_A$ and $c_S$ are the area specific heats of the atmosphere and surface layers, respectively. Eqs. (1) and (2) of the text refer to the case in which $E_A$ and $E_S$ are constants. We have assumed, as in other EBM treatments (*e. g.*, [15]), that in the region of temperatures concerned, temperature-independent specific heats exist such that $dE_A = c_A dT_A$ and $dE_S = c_S dT_S$.

Assuming that the temperatures make small departures from their steady-state values $T_{A0}$ and $T_{S0}$, such that $T_A = T_{A0} + \Delta T_A$ and $T_S = T_{S0} + \Delta T_S$, we have, by Taylor expansion, assuming that $\Delta T_A$ and $\Delta T_S$ are small quantities:

$$F_A = F_{A0} + q_A \Delta T_A \text{ and } F_S = F_{S0} + q_S \Delta T_S, \tag{A3,4}$$

where $F_{A0}$ and $F_{S0}$ are the Stefan-Boltzmann fluxes $\sigma T_{A0}^4$ and $\sigma T_{S0}^4$, and $q_A = 4\sigma T_{A0}^3$, $q_S = 4\sigma T_{S0}^3$. In addition to these explicitly temperature-dependent quantities, nearly all the parameters in Eqs. (A1,2) may well depend on $T_A$ or $T_S$ or both. The equations are therefore generally nonlinear. However, for small anomalies the Taylor expansion method may be extended to the parameters as well as the $F$'s. We therefore write

$$A(t) = A_0 + A_1 \Delta T_A(t), \quad B(t) = B_0 + B_1 \Delta T_S(t), \quad S_{NR}(t) = S_{NR0} + q_{NR} \Delta T_S(t), \tag{A5,6,7}$$

where the new parameters $A_1$, $B_1$ and $q_{NR}$ are adjustable unknowns or may be estimated from consideration of Taylor expansion coefficients. For example, $q_{NR} = \partial S_{NR}/\partial T_S$, with other variables constrained as required by the context.

In practice, we are interested in the effect of imposed variations of the solar flux and of the composition of the atmosphere. In the former case we have

$$S(t) = S_0 + \Delta S(t). \tag{A8}$$

In this paper we attribute the effect of Pinatubo's eruption to extrinsic changes in $A_0$ and $B_0$, ignoring the aerosol's effect on $\varepsilon$. In other applications, where changes of greenhouse gas concentrations are of interest, there would be some small effects on $A$ and $B$, we assume these to be negligible at this modeling stage. For $\varepsilon$, we consider the possibility of both an externally imposed change $\Delta\varepsilon_0(t)$ and a system-induced temperature dependence, so that

$$\varepsilon(t) = \varepsilon_0 + \Delta\varepsilon(t), \tag{A9}$$

where

$$\Delta\varepsilon(t) = \Delta\varepsilon_0(t) + \varepsilon_{1A}\Delta T_A(t) + \varepsilon_{1S}\Delta T_S(t). \tag{A10}$$

Finally, the imbalance parameter $b$ must be considered. Following a suggestion by R. Henry [21], we assume that $b$ depends primarily on $\varepsilon$ and write

$$b = b_0 + b_1\Delta\varepsilon, \tag{A11}$$

which, through Eq. (A10), gives $b$ an implicit temperature dependence as well as a dependence on $\Delta\varepsilon_0(t)$.

**b. Construction of the working equations.** Upon substituting Eqs. (A3-11) into Eqs. (A1, 2) one obtains terms containing "$\Delta$" factors to the zeroth, first, and second power. Only those with first power "$\Delta$" factors are retained. Those with no "$\Delta$" factors comprise the expression for the steady state of the system (because the left hand side is zero). Those with two or more "$\Delta$" factors are neglected in order to be consistent with the neglect of the higher terms in $F_A$ and $F_S$, in keeping with the general philosophy of linearization. These assumptions must of course be revisited if the solutions appear to be driven out of the linear regime.

In what follows, for economy and without likelihood of confusion, we drop the subscripts "0" from $b_0$, $\varepsilon_0$, $A_0$, $B_0$, $F_{A0}$, $F_{S0}$, and $S_0$. The working equations become

$$c_A \frac{d\Delta T_A(t)}{dt} = -K_{AA}\Delta T_A(t) - K_{AS}\Delta T_S(t) + \Delta n_A(t) \tag{A12}$$

$$c_S \frac{d\Delta T_S(t)}{dt} = -K_{SA}\Delta T_A(t) - K_{SS}\Delta T_S(t) + \Delta n_S(t) - Q \tag{A13}$$

where the $K$ coefficients are defined by

$$\begin{aligned} K_{AA} &= (1+b)\varepsilon q_A(1-f_{AA}), \\ K_{AS} &= -\varepsilon q_S(1-f_{AS}), \\ K_{SA} &= -b\varepsilon q_A(1-f_{SA}), \\ K_{SS} &= q_S(1-f_{SS}), \end{aligned} \tag{A14a-d}$$

and the driving terms are

$$\Delta n_A(t) = [F_S - (1+b+b_1\varepsilon)F_A]\Delta\varepsilon_0(t) + A\Delta S(t) + S\Delta A(t), \tag{A15}$$

$$\Delta n_S(t) = (b+b_1\varepsilon)F_A\Delta\varepsilon_0(t) + B\Delta S(t) + S\Delta B(t). \tag{A16}$$

As pointed out in the text, the coupling term $Q$ is proportional to $\Delta T_S$ and is easily incorporated into the linearized equations.

In the $K$'s, all factors of the form $(1-f)$ become unity when there is no parameter temperature dependence, i. e., when there is no feedback. A factor $f_{XY}$ may be thought of as the feedback on the energy balance of box $X$ due to a change in temperature $T_Y$. One may have the impression that by choosing arbitrary values of the $f$'s nearly anything could be predicted with these equations. This is true, except that the $f$'s are not in fact arbitrary. They are rather complicated combinations of the physical parameters introduced above:

$$f_{AA} = \frac{[F_S - (1 + b + b_1\varepsilon)F_A]\varepsilon_{1A} + SA_1}{(1+b)\varepsilon q_A}, \tag{A17a}$$

$$f_{AS} = \frac{[(1 + b + b_1\varepsilon)F_A - F_S]\varepsilon_{1S} - q_{NR}}{\varepsilon q_S}, \tag{A17b}$$

$$f_{SA} = -\frac{(b + b_1\varepsilon)F_A\varepsilon_{1A} + SA_1}{b\varepsilon q_A}, \tag{A17c}$$

$$f_{SS} = -\frac{(b + b_1\varepsilon)F_A\varepsilon_{1S} - q_{NR} + SB_1}{q_S}, \tag{A17d}$$

These feedbacks are expected to lie in the range $\{-\infty < f < 1\}$, which excludes oscillatory behavior.

In some contexts the individual feedback parameters are usefully combined into the following overall effective feedback, particularly in the case of the climate sensitivity. An expression such as Eq. (10) of the text can be shown to acquire an additional factor $1/(1 - f_{eff}^S)$, where

$$f_{eff}^S = 1 - \frac{1 - \dfrac{(1+b)(f_{AA} + f_{SS} - f_{AA}f_{SS}) - b\varepsilon_0(f_{AS} + f_{SA} - f_{AS}f_{SA})}{1 + b - b\varepsilon_0}}{1 - \dfrac{f_{AA}(1+b)B + f_{SA}bA}{(1+b)B + bA}}, \tag{A18}$$

## Appendix B. Solutions

The Laplace-transformed solutions, Eqs. (11,12), may be converted readily into linear combinations of exponential decays in the case $\kappa = 0$. The time constants of interest are the inverse of the eigenvalues of the kinetic matrix

$$\begin{pmatrix} K_{AA}/c_A & K_{AS}/c_A \\ K_{SA}/c_S & K_{SS}/c_S \end{pmatrix} = \begin{pmatrix} k_{AA} & k_{AS} \\ k_{SA} & k_{SS} \end{pmatrix}, \tag{B1}$$

having the values

$$\begin{pmatrix} \lambda_1 \\ \lambda_2 \end{pmatrix} = \begin{pmatrix} (k_{AA} + k_{SS})/2 - \sqrt{(k_{AA} - k_{SS})^2/4 + k_{AS}k_{SA}} \\ (k_{AA} + k_{SS})/2 + \sqrt{(k_{AA} - k_{SS})^2/4 + k_{AS}k_{SA}} \end{pmatrix}. \tag{B2}$$

The resulting solutions, after quite a bit of tedious linear algebra, may be expressed as

$$u_A(t) = \frac{-0.439 A}{(\lambda_2 - \lambda_1)t_V} \left[ \left(\frac{\phi_A}{c_A}\right) \Omega_2(t) - k_{AS} \left(\frac{\phi_S}{c_S}\right) \Omega_1(t) \right] \tag{B3}$$

and

$$u_S(t) = \frac{-0.439 A}{(\lambda_2 - \lambda_1)t_V} \left[ \left(\frac{\phi_S}{c_S}\right) \Omega_3(t) - k_{SA} \left(\frac{\phi_A}{c_A}\right) \Omega_1(t) \right], \tag{B4}$$

where

$$\Omega_1(t) = \frac{\exp(-\lambda_1 t)}{(k_V - \lambda_1)^2} - \frac{\exp(-\lambda_2 t)}{(k_V - \lambda_2)^2} + \left[ \frac{1 + (k_V - \lambda_2)t}{(k_V - \lambda_2)^2} - \frac{1 + (k_V - \lambda_1)t}{(k_V - \lambda_1)^2} \right] \exp(-k_V t), \tag{B5}$$

$$\Omega_2(t) = \frac{k_{SS} - \lambda_1}{(k_V - \lambda_1)^2} \exp(-\lambda_1 t) - \frac{k_{SS} - \lambda_2}{(k_V - \lambda_2)^2} \exp(-\lambda_2 t) + \ldots$$

$$+ \left[ \frac{(k_{SS} - \lambda_2)[1 + (k_V - \lambda_2)t]}{(k_V - \lambda_2)^2} - \frac{(k_{SS} - \lambda_1)[1 + (k_V - \lambda_1)t]}{(k_V - \lambda_1)^2} \right] \exp(-k_V t), \tag{B6}$$

and

$$\Omega_3(t) = \frac{k_{AA} - \lambda_1}{(k_V - \lambda_1)^2} \exp(-\lambda_1 t) - \frac{k_{AA} - \lambda_2}{(k_V - \lambda_2)^2} \exp(-\lambda_2 t) + \ldots$$

$$+\left[\frac{(k_{AA}-\lambda_2)[1+(k_V-\lambda_2)t]}{(k_V-\lambda_2)^2}-\frac{(k_{AA}-\lambda_1)[1+(k_V-\lambda_1)t]}{(k_V-\lambda_1)^2}\right]\exp(-k_V t). \quad (B7)$$

In the case $\phi_S = 1$, $\phi_A = 0$, and $k_{AS} = k_{SA} = 0$ (reducing the problem effectively to one layer), solution (B4) reduces to the one obtained by Douglass and Knox [5], noting the additional correspondences $BS_0 = A$, $k_{SS} = 1/\tau = 1/(c_S\lambda)$.

***Including coupling to the ocean.*** Eqs. (9,10) may be solved directly for the Laplace-transformed anomalies:

$$\tilde{u}_A(p) = \frac{-0.439 A}{t_V} \cdot \frac{1}{(p+k_V)^2} \cdot \frac{\left(p+k_{SS}+\kappa' p^{1/2}\right)(\phi_A/c_A)-k_{AS}(\phi_S/c_S)}{(p+k_{AA})\left(p+k_{SS}+\kappa' p^{1/2}\right)-k_{AS}k_{SA}} \quad (B8)$$

and

$$\tilde{u}_S(p) = \frac{-0.439 A}{t_V} \cdot \frac{1}{(p+k_V)^2} \cdot \frac{(p+k_{AA})(\phi_S/c_S)-k_{SA}(\phi_A/c_A)}{(p+k_{AA})\left(p+k_{SS}+\kappa' p^{1/2}\right)-k_{AS}k_{SA}}. \quad (B9)$$

Preliminary work has shown that these transformed expressions may be inverted successfully in the case $\kappa > 0$ by using the same contour as that use in ref. [13] and that this two-box version reproduces the one-box ocean-delayed solution found there.

## Appendix C.  The meaning of $\lambda$ and $\tau$ in a two-box model

In a one-box model, the dynamical equation for the temperature anomaly is usually written in the form

$$hc\frac{d\Delta T(t)}{dt} = -\frac{\Delta T(t)}{\lambda} + \Delta F(t), \tag{C1}$$

where $hc$ is the effective heat capacity of the box and where $\lambda$ can be identified with the asymptotic ("equilibrium") temperature shift associated with a step-function forcing $\Delta F_0$:

$$\Delta T(\infty) = \lambda \Delta F_0. \tag{C1}$$

$\lambda$ remains a parameter under different forcings and may be used to fit the data.  However, in a model any having more than one box, $\lambda$ is a derived parameter, *i. e.*, it must be defined and extracted from the theory.  Since the one-box $\Delta T$ is used to describe the surface temperature, in the two-box case it seems most reasonable to compute a quantity $\lambda_S$ as the response $\Delta T_S(\infty)$ to a step-function forcing.  According to Eqs. (A12,13),

$$K_{AA}\Delta T_A(\infty) + K_{AS}\Delta T_S(\infty) = \phi_A \Delta F_0, \tag{C3}$$

$$K_{SA}\Delta T_A(\infty) + K_{SS}\Delta T_S(\infty) = \phi_S \Delta F_0, \tag{C4}$$

from which we find

$$\lambda_S = \frac{b(1-f_{SA})\phi_A + (1+b)(1-f_{AA})\phi_S}{(1+b)(1-f_{AA})(1-f_{SS}) - b\varepsilon(1-f_{AS})(1-f_{SA})} \cdot \frac{1}{q_S}. \tag{C5}$$

The one-box task of fitting data by "adjusting $\lambda$" is seen to have ballooned into selecting all the feedbacks.  This is what one is doing, effectively, when adjusting $\lambda$ in the one-box model, where there is only one feedback parameter.

When there is no feedback, Eq. (C5) yields

$$\lambda_{S0} = \frac{b\phi_A + (1+b)\phi_S}{1+b-b\varepsilon} \cdot \frac{1}{q_S} = \frac{bA + (1+b)B}{(1-\alpha)(1+b-b\varepsilon)} \cdot \frac{1}{q_S}. \tag{C6}$$

Eq. (C6) is algebraically equivalent to text Eq. (13).

There is no easy connection between the two-box lifetime parameters and the one-box $\tau = hc\lambda$, since the two-box $\Delta T_S$ depends on two relaxation times. As a representative value of $\tau$ we may use the inverse of the smaller eigenvalue, $\lambda_1$ of Eq.(B2), which will dominate the behavior of both $\Delta T_A$ and $\Delta T_S$ except at short times.

**References**

1. K. P. Shine, Y. Fouquart, V. Ramaswamy, S. Solomon, and J. Srinivasan, Radiative Forcing. in *Climate Change 1994*, edited by J. T. Houghton *et al.*, Cambridge Univ. Press, Cambridge, 1995, pp. 162-204.

2. J. Hansen, A. Lacis, R. Ruedy, and M. Sato (1992), Potential climate impact of Mount Pinatubo eruption, Geophys. Res. Lett. **19**, 215-218

3. R. S. Lindzen (1994), Climate dynamics and global change, Ann. Rev. Fluid Mech. **26**, 353-378

4. R. S. Lindzen (1995), Constraining possibilities versus signal detection in Martinson-D-G, Natural Climate Variability on Decade-to-Century Time Scales (National Academy Press, Washington, DC), pp. 182-186

5. D. H. Douglass and R. S. Knox, "Climate forcing by the volcanic eruption of Mount Pinatubo," Geophys. Res. Lett. **32** (5), L05710, doi:10.1029/2004GL022119 (2005). Revised version archived as http://arXiv.org/abs/physics/0509166.

6. J. Hansen, M. Sato, and R. Ruedy (1997), Radiative forcing and climate response, J. Geophys. Res. **102**, 6831-6864

7. Hansen, J., and 27 others (2002), Climate forcings in Goddard Institute for Space Studies SI2000 simulations, J. Geophys. Res., **107**(D18), 4347, doi:10.1029/2001JD001143.

8. C. M. Ammann, G. A. Meehl, and W. W. Washington (2003), A monthly and latitudinally varying volcanic forcing data set in simulations of 20th century climate, Geophys. Res. Lett., 30(12), 1657, doi:10.1029/2003GL016875

9. T. M. L. Wigley, C. M. Ammann, and B. D. Santer (2005), Using the Mount Pinatubo volcanic eruption to determine climate sensitivity: Comments on "Climate forcing by the volcanic eruption of Mount Pinatubo" by David H.

Douglass and Robert S. Knox, Geophys. Res. Lett. **32**, L20711, doi:10.1029/2005GL023312

10. A. Robock (2005), Using the Mount Pinatubo volcanic eruption to determine climate sensitivity: Comments on "Climate forcing by the volcanic eruption of Mount Pinatubo" by David H. Douglass and Robert S. Knox, Geophys. Res. Lett. L20709, doi:10.1029/2005GL023287

11. D. H. Douglass and R. S. Knox (2005), "Reply to comment by T. M. Wigley et al. on 'Climate forcing by the volcanic eruption of Mount Pinatubo'," Geophys. Res. Lett. **32**, L20710, doi:10.1029/2005GL023695

12. D. H. Douglass and R. S. Knox (2005), "Reply to comment by A. Robock on 'Climate forcing by the volcanic eruption of Mount Pinatubo'," Geophys. Res. Lett. **32**, L20712, doi:10.1029/2005GL023829

13. D. H. Douglass, R. S. Knox, B. D. Pearson, and A. Clark, Jr. (2006), Thermocline flux exchange during the Pinatubo event, Geophys. Res. Lett. **33**, L19711, doi:10.1029/2006GL026355

14. R. S. Knox (2004), Non-radiative energy flow in elementary climate models, Phys. Lett. A **329**, 250-256

15. K. M. Shell and R. C. J. Somerville (2005), A generalized energy balance climate model with parameterized dynamics and diabatic heating, J. Climate **18**, 1753-1772

16. J. R. Christy, R. W. Spencer, and W. D. Braswell (2000), MSU tropospheric temperatures: dataset construction and radiosonde comparisons, *J. Atmos. Oceanic Tech., 17*, 1153-1170. Updates available at http://vortex.nsstc.uah.edu/data/msu/t2lt/.

17. This sensitivity is the response to a "direct" forcing at the top of the atmosphere due to a variation in the solar flux. It differs slightly from the standard IPCC sensitivity to "radiative forcing" in two ways. First, "radiative forcing" is defined by a two-stage process discussed in [1] generally and in [14] as it relates to the present model. For our purposes the difference in $\lambda$ is negligible. Second, the factor $1/(1-\gamma)$ is not normally included in standard treatments, probably because it is considered a feedback factor. Without this factor, one obtains the standard "Stefan-Boltzmann" result $\Delta T_S/\Delta F = 0.30$ °C/(w/m$^2$). With the factor included, 0.30 changes to 0.36.

18. R. S. Knox (1999), Physical aspects of the greenhouse effect and global warming, Amer. J. Phys. **67**, 1227-1238

19. J. R. Barker and M. H. Ross (1999), An introduction to global warming, Amer. J. Phys. **67**, 1216-1226

20. D. L. Hartmann (1994), *Global Physical Climatology* (Academic Press, San Diego *et al.*), pages 114 and 375.

21. R. L. Henry (2008), private communication.

**Table 1.** Static fitting parameters defining the steady-state problem, as discussed in text. See also Figure 1.

|  | Units | Static set 1* | Static set 2* |
|---|---|---|---|
| $T_S$ | K | **288** | **288** |
| $T_A$ | K | 243 | 249 |
| $\alpha = 1 - (A + B)$ |  | 0.30 | 0.313 |
| $b$ |  | 1.65 | 1.66 |
| $\varepsilon$ |  | 0.786 | 0.882 |
| $S_{NR}$ | W/m$^2$ | 100 | 102 |
| $A$ (atm. absorp. fraction) |  | **0.02** | **0.20** |
| $B$ (sfc. absorp. fraction) |  | **0.68** | **0.50** |
| $S_0$ | W/m$^2$ | **342** | **342** |
| $q_S$ | W/m$^2$/K | *5.42* | *5.42* |
| $q_A$ | W/m$^2$/K | *3.23* | *3.52* |
| $\phi_A$ |  | 0.029 | 0.285 |
| $\phi_S$ |  | 0.971 | 0.715 |

\* Quantities in bold fixed; others chosen to satisfy the criteria of reasonability [14,18]. Derived quantities needed for the dynamical equations are shown in italics

**Table 2.** Dynamical fitting parameters, as discussed in text. The derived quantities $\lambda_S$ and $\tau$ are defined and discussed in Appendix C.

|  | Units | Static set 1 | Static set 2 |
|---|---|---|---|
| $f_{AA}$ |  | 0.0 | 0.0 |
| $f_{AS}$ |  | –0.31 | –0.34 |
| $f_{SS}$ |  | –0.62 | –0.64 |
| $f_{SA}$ |  | 0.0 | 0.0 |
| $f_{eff}$ |  | –0.93 | –1.02 |
| $h_m$ | m | 18 | 21 |
| $\lambda_S$ | °C/(W/m$^2$) | 0.19 | 0.18 |
| $\tau$ | months | 5.8 | 7.2 |
| $A$ (forcing parameter) |  | 21 | 21 |

Figure 1. The two-layer box model and its principal fluxes, with coupling to a box representing the deeper ocean ("thermocline"). For full description of the parameters in the atmosphere and surfaces boxes, see text following Eqs. (1) and (2). Edited version of Fig. 1 of Knox [14].

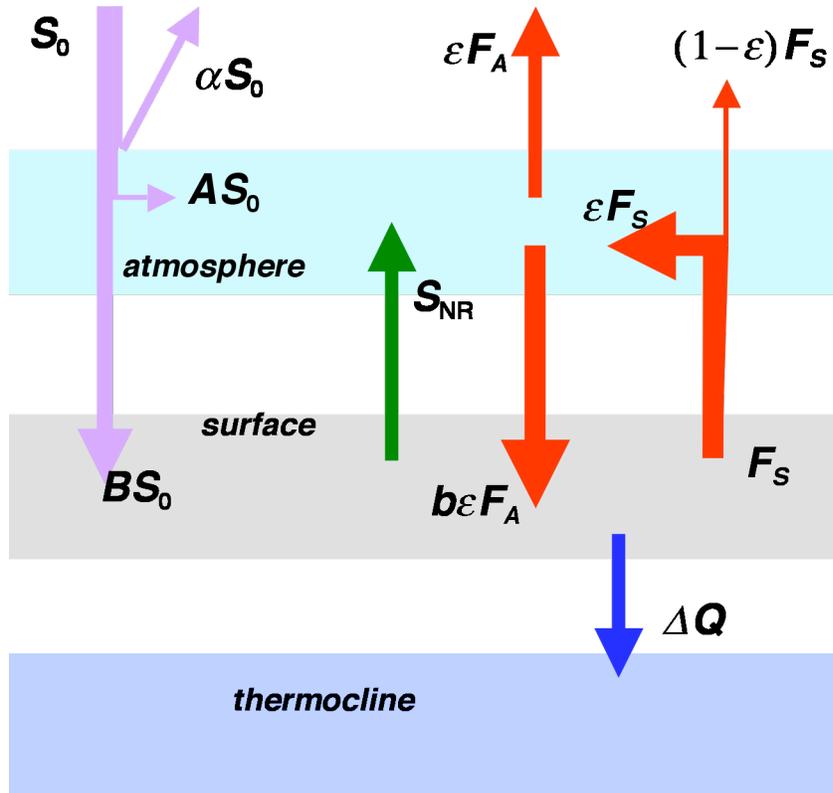

Figure 2. Time dependent atmosphere and surface temperature anomalies induced by the Mt. Pinatubo forcing using parameter set 1 (forcing mainly felt at the surface layer). Circles are the TLTm data set. Here a best fit of the atmosphere layer solution ($u_B$) to the data (blue curve) has been found, with feedbacks constrained to zero and by varying the effective surface layer parameter $h_m$.

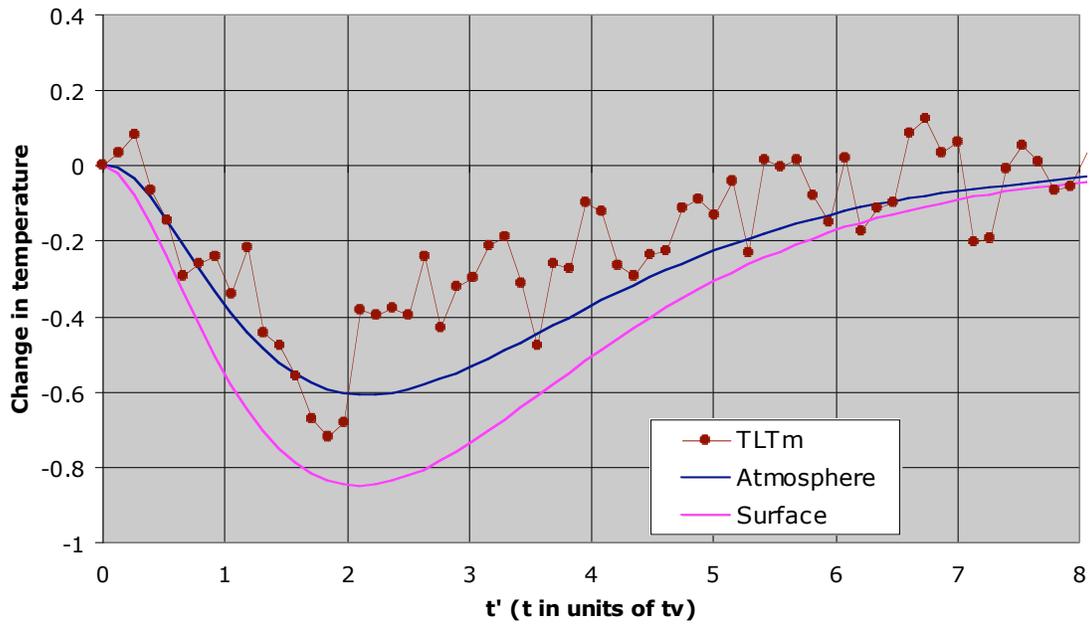

Figure 3. Plot of the time dependent atmosphere and surface temperature anomalies induced by the Mt. Pinatubo forcing using parameter set 1 (forcing mainly felt at the surface layer). Circles are the TLTm data set. Here a best fit has been found with the atmosphere layer solution ($u_B$) to the data (blue curve). Results with parameter set 2, not shown, with forcing felt in both layers, are identical.

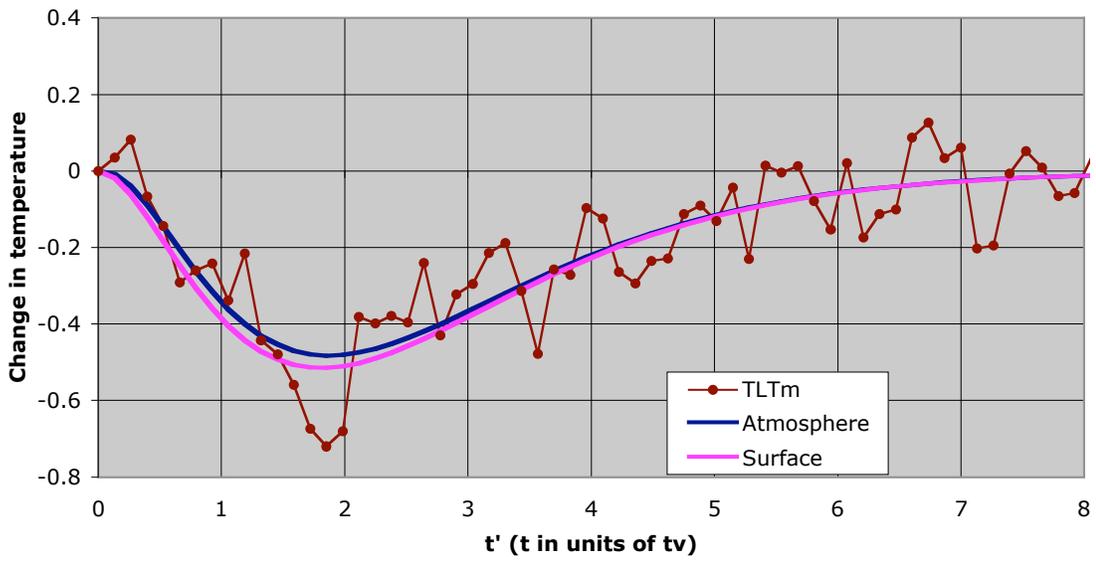

Figure 4. Plot of the time dependent atmosphere and surface temperature anomalies induced by the Mt. Pinatubo forcing using parameter set 1 (forcing mainly felt at the surface layer). Circles are the TLTm data set. Here a best fit has been found with the surface-layer solution ($u_S$) (pink curve) to the data.

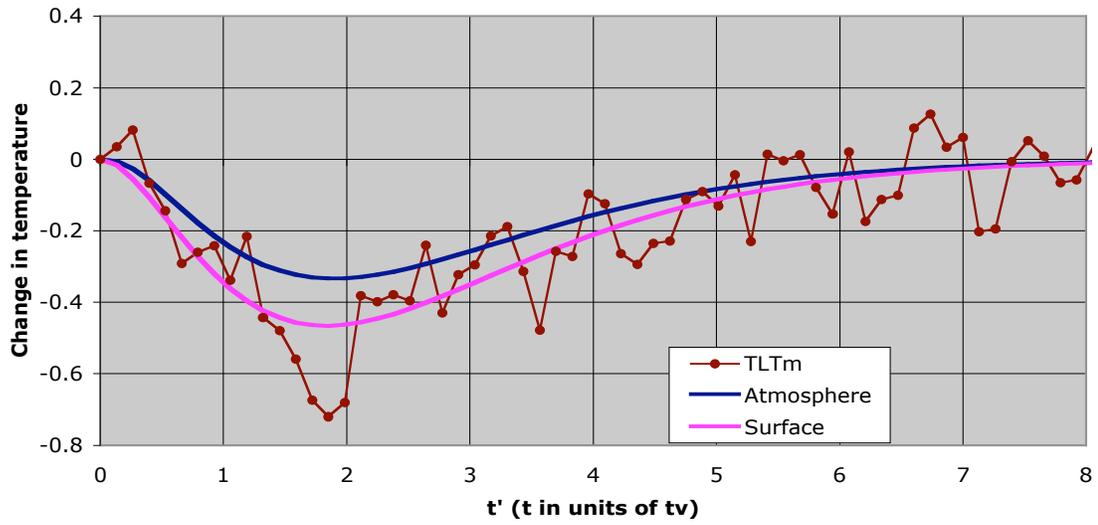